\documentclass[aps,prl,twocolumn,showpacs]{revtex4}

\usepackage{amsmath} 
\usepackage{amssymb} 
\usepackage{epsfig}
\usepackage{xspace}
\usepackage{graphicx}

\newcommand{\p}{\partial} 
\newcommand{\gk}{\Gamma_k}
\newcommand{\anz}{{\it ansatz}\xspace}
\newcommand{\half}{\displaystyle \frac{1}{2}}
\newcommand{\trho}{\tilde{\rho}}
\newcommand{\pms}{{\hbox{\tiny PMS}}}

\def\nbZ{{\mathchoice {\hbox{$\sf\textstyle Z\kern-0.4em Z$}}
{\hbox{$\sf\textstyle Z\kern-0.4em Z$}}
{\hbox{$\sf\scriptstyle Z\kern-0.3em Z$}}
{\hbox{$\sf\scriptscriptstyle Z\kern-0.2em Z$}}}}

\begin{document}

\title{Optimization of field-dependent nonperturbative renormalization
 group flows}

\author{L\'eonie Canet}
\affiliation{Laboratoire de Physique Th\'eorique et Hautes \'Energies,
Universit\'{e}s Paris VI -- Pierre et Marie Curie, Paris VII -- Denis Diderot, 2 place Jussieu, 75251 Paris cedex 05, France}

\begin{abstract}
We investigate the influence of the momentum cutoff function on 
  the  field-dependent nonperturbative 
 renormalization group  flows for the three-dimensional Ising model,
 up to the second order  of the derivative expansion. We show
  that, even when dealing with the full functional dependence of the
 renormalization functions,  the accuracy of the critical exponents 
 can be simply optimized, through the principle of minimal sensitivity,
  which yields $\nu = 0.628$ and $\eta = 0.044$. 
\end{abstract}


\date{\today}

\maketitle

Many  phenomena in high-energy as well as in statistical physics cannot be tackled using
 perturbative methods, 
   either because  they come under the strong coupling regime of their field theory 
  or because they appear genuinely nonperturbative like, for instance, confinement in QCD
  or phase transitions induced  by topological defects.
  The  Wilsonian renormalization group concept,
  resting on progressive integration of high-energy fluctuations
  \cite{wilson74},  has rooted
   the nonperturbative renormalization group (NPRG) formalism. 
  The latter has arousen a renewed attention, since the idea emerged 
  \cite{wetterich93c,ellwanger94c,morris94a}
  to apply this procedure    
   to  the Legendre transform $\Gamma$ of the free energy --- the Gibbs
 free energy --- rather than to the Hamiltonian.    
  This method has 
  allowed to broach notoriously difficult problems,  such as  the  abelian   Higgs  model   relevant  for
 superconductivity \cite{bergerhoff96},   the   phase   diagram   of
 He$_3$ \cite{kindermann01},           the
 Gross-Neveu  model  in three  dimensions \cite{rosa01,hofling02}, frustrated  magnets \cite{tissier03} or reaction-diffusion
 processes \cite{canet04,canet04a} (see
 \cite{berges02} and \cite{bagnuls01} for  reviews).

 The NPRG formalism consists in building a sequence of running effective average actions $\Gamma_k$, 
  which are infrared regulated by a momentum cutoff and thus only include fluctuations with momenta larger than the  
 running  scale $k$.
 $\Gamma_k$  continuously
 interpolates between the microscopic action $\Gamma_{k=\Lambda} = S$ and  the 
 Gibbs free energy $\Gamma_{k=0} =\Gamma$. Its flow   
  with $k$ is governed by an exact equation \cite{tetradis94}:
\begin{equation}
\p_k \gk (\psi)=\half \hbox{Tr} \left\{ [\gk^{(2)}+{R}_k]^{-1}\,\p_k {R}_k\right\},
\label{flowgam}
\end{equation}
where ${\Gamma}_k^{(2)}$ is the full inverse propagator and  $R_k$ is the
infrared cutoff
 function 
  \footnote{In Eq. (\ref{flowgam}), Tr stands for trace over internal indices and integration
 over  internal momenta.}.  

 Though Eq. (\ref{flowgam}) is exact, hence preserving all nonperturbative features of the model,
  it is functional and,
 obviously, cannot be solved exactly. Any practical calculation 
  requires to truncate  $\gk$, most commonly through
  a derivative expansion \cite{tetradis94}.
   Furthermore, when the number of invariants or of renormalization
  functions  grows, one often resorts to
   an additional truncation  to simplify the
 numerical task, which consists in field expanding 
  the different renormalization functions (\cite{tetradis94} and see below).

  Naturally, these expansions raise questions as for their convergence
 and accuracy,
  since these properties entirely condition the  reliability 
  of the method, and many works have thus been devoted to their study  
  \cite{papenbrock95,morris99,zumbach94b,aoki98,liao00,litim02,ball95,comellas97,mazza01}.
   These issues have appeared intimately related to that of the influence
  of the cutoff function  $R_k$. Although
  the exact solution $\Gamma = \lim_{k\to0} \gk$ does not depend on $R_k$,
   any truncation breaks this invariance. This suggests that the
 choice 
 of the cutoff may  be optimized, and
   various criteria have been proposed,
   such as the optimization of the rapidity of 
  the  convergence of the field expansion \cite{halliday80,liao00}, 
  the maximization of   the ``gap'' of the propagator \cite{litim00,litim01,litim01c,litim02}  or
  the principle of minimal sensitivity (PMS) \cite{stevenson81,canet03a,canet03b}.
  All these analyses rely on the field expansion of the renormalization functions.

  However,  a field expansion cannot always be performed. First, it
  is meaningless in dimensions where the field canonical dimension
  vanishes, since all  field powers then become equally relevant. More importantly, some phenomena
  require, in principle,  a functional description,
  for instance if the effective potential develops non-analycities, 
 which can  occur for some disordered systems \cite{fisher86,wegner73,ledoussal04} 
  and continuous growth models \cite{ledoussal03}.
  Since the optimization is crucial to control  approximations,
  one has to dispose of an efficient  procedure 
  to deal with the field-dependent renormalization functions.
  However, this had so far never been investigated as it involves a sizeable
  numerical task. In this article, we show that, even without
 truncating the field-dependence, 
 the PMS allows to simply optimize the choice of $R_k$,  and  we provide
   the corresponding optimal critical exponents for the three-dimensional Ising model,
     up to the second order of the derivative expansion.\\

  Let us first draw the framework of this calculation. The derivative expansion  physically rests 
 on the assumption that the long-distance physics --- corresponding
 to the low-energy modes ($q\to0$) --- 
 is well captured  by the lowest order derivative terms.
 Hence, for the Ising model, the \anz of $\gk$ at order $\p^2$ writes \cite{tetradis94}:
\begin{equation}
 \gk = \int d^d x \, \Big\{ U_k(\rho) + \half Z_k(\rho)\,(\nabla \psi)^2  \Big\},
\label{anz}
\end{equation}
 where $\rho=\psi^2/2$ is the $\nbZ_2$ invariant. The running potential $U_k(\rho)$ describes 
 the physics associated with uniform field configurations while the field renormalization function
 $Z_k(\rho)$ renders the effect of slowly varying fields.
  The flow equations of the running functions $U_k$ and $Z_k$ follow
  from Eq. (\ref{flowgam}),
 respectively through its evaluation at a uniform field configuration
 and through isolating the $q^2$ dependence of its second
 functional derivative with respect to  $\psi(q)$ and $\psi(-q)$ \cite{tetradis94}.  
 The running anomalous dimension is then defined by \cite{berges02}:
\begin{equation}
\eta_k = - \p_t \, \ln Z_{k,0}, 
\label{defeta}
\end{equation}
where $t=\ln(k/\Lambda)$   and $Z_{k,0}\equiv Z_k(\rho_0)$
 is the field renormalization at the running minimum $\rho_0(k)$ of the potential. 
 We introduce the dimensionless and renormalized variables $\trho = Z_{k,0}\,k^{2-d}\,\rho$, 
 $u_k = k^{-d}\,U_k$  and $z_k =
Z_{k,0}^{-1}\,Z_k$, which are more convenient for the search of fixed points.

  The field expansion 
   would then consist in expanding $u_k$ and $z_k$ in powers of the 
 invariant $\trho$ around the running minimum $\trho_0(k)$ of the potential
 --- the latter point conveying 
 nice convergence properties \cite{liao00,litim02} --- which  would write
  for a generic function $h_k$:
\begin{equation}
 h_k(\trho) = \sum_{n=0}^{p_h}\, \frac{1}{n!}\, h_{k,n}\,(\trho -\trho_0)^n.
\label{fieldexp}
\end{equation}
 
 We here do not proceed to such an expansion. This expression is  given 
  for completeness since we shall compare, in the following, our results with those
 obtained in \cite{canet03a}  ensuing from a field truncation of  
 the  renormalization functions $u_k$ and $z_k$
  to the tenth and ninth order respectively 
 --- that is  $p_u = 10$ and $p_z =9$ in Eq. (\ref{fieldexp}).

 To determine the critical exponents  of the Ising model,
  we numerically integrate the flow equations $\p_t u_k(\trho)$ and 
 $\p_t z_k(\trho)$ from an initial microscopic scale $\Lambda$ --- at 
 which the potential is quartic $u_{\Lambda}(\trho) = \lambda/2\,(\trho -\trho_0)^2$ and
  the field renormalization 
 is constant $z_{\Lambda}(\trho)=1$ --- to the physical scale $k\to 0$. For large values of the
 initial  parameter $\trho_0(\Lambda)$, the running 
 minimum $\trho_0(k)$ of the potential flows to infinity  such that  the dimensionful renormalized minimum ---
 the magnetization ---  acquires a finite positive value (broken phase). 
 For small initial $\trho_0(\Lambda)$, the running minimum $\trho_0(k)$, and thus the magnetization, vanishes at a finite scale $t_s$  (symmetric phase).
  In between, there exists a
 critical initial parameter ${\trho_0}^{cr}(\Lambda)$
 for which the running potential reaches a fixed point \cite{seide99}. The critical exponent $\eta$
  is then given  by the fixed point
  value $\eta^*$ of the running anomalous dimension  $\eta_k$ \cite{seide99,berges02}. 
 The critical exponent $\nu$ --- describing the divergence of the
 correlation length near criticality ---  is related to the negative eigenvalue 
  of the stability matrix, corresponding to the linearized flows of
 the renormalization functions on the field mesh, in the vicinity of the fixed point.

 Setting $z_k(\trho)=1$ for all $k$, {\it i.e.} neglecting the field renormalization constitutes the Local Potential
 Approximation (LPA). Within the LPA, the anomalous dimension $\eta$ remains zero. 
 Including a running field renormalization coefficient $Z_{k,0}$, independent of the field, 
  allows one to refine the LPA by providing a non-trivial --- although rough --- 
 determination of $\eta$. This approximation is referred
 to as First Order Approximation (FOA) in the following.
  The next step then consists in incorporating the full field-dependence
 of the renormalization function $z_k(\trho)$, which is here called  Second Order Approximation
 (and denoted SOA). 

 We successively study these three levels of approximations, and 
  analyze for each the influence of $R_k$. We consider
  an exponential cutoff --- which achieves an efficient separation of the
 low- and high-energy modes \cite{tetradis94} ---  parametrized by a free amplitude $\alpha$:
\begin{equation}
\displaystyle{r(y)=\alpha\, \frac{1}{e^y-1}},
\end{equation}
where $r(y) = R_k(q^2) /(Z_{k,0}\,q^2)$ and $y = q^2/k^2$.

\begin{figure}[ht]
\begin{center}
\includegraphics[width=48mm,height=69mm,angle=-90]{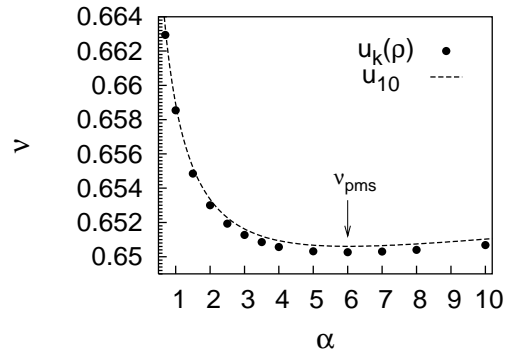}
\caption{Variations of $\nu$ with $\alpha$ at the LPA. The dashed line represents
 the results obtained with a tenth order truncated potential \cite{canet03a}, dots those
 obtained keeping the full functional dependence of $u_k(\trho)$ (present work).}
\label{pmsd0}
\end{center}
\end{figure}

 For each parameter $\alpha$, the initial value of  $\trho_0(\Lambda)$ is fine-tuned to reach
  the critical regime and the associated exponents are computed. 
  The optimal $\alpha$ is then determined through the PMS, that is
 for a specific exponent $\chi$,  the parameter
 $\alpha_\pms$ for which  $\chi$ is stationary is selected:
\begin{equation}
\left.\frac{d\, \chi(\alpha)}{d\, \alpha}\right|_{\alpha_\pms}=0,
\label{pms}
\end{equation}
and $\chi_\pms \equiv \chi(\alpha_\pms)$ is identified as the optimized exponent.
 The hope is that imposing a constraint satisfied by
 $\chi$ in the exact theory  improves the
 approximate determination  of this quantity. The
 obvious drawback of this criterion is
that Eq. (\ref{pms}) may exhibit many solutions. But in fact,
  this can be easily circumvented 
 by complementing the PMS with a few  basic rules
  that allow to simply discriminate between   multiple solutions \cite{canet03a,canet03b}. \\


We can now set out the results, which are collected in Table \ref{tab}
 along with the best theoretical determinations. 
 Figure \ref{pmsd0} displays the variations of $\nu$ with the parameter
 $\alpha$ within the LPA,
     together with
  the analogous curve ensuing from the field truncation of the
 potential (at $p_u =10$) \cite{canet03a}. 
  Both $\nu(\alpha)$ curves --- with and without field truncation --- appear to coincide within a few tenths of percents. This agreement confirms the convergence of the field expansion
 and thus validates the former analysis \cite{canet03a}.
 Moreover,  the function  $\nu(\alpha)$  exhibits a unique extremum, which hence embodies
 the stationary solution of the PMS. The corresponding optimal exponent
 is $\nu_\pms = 0.6503$.

 Let us emphasize that within the LPA,   $\nu(\alpha)$ always 
 overvalues the ``expected'' exponent $\nu=0.6304(13)$ (according
 to  the best theoretical 
 estimates),
   and that $\nu_\pms$ 
 corresponds to the minimum. Thus, as already highlighted in \cite{canet03a}, the PMS  selects the
  most accurate exponent achievable within a given approximation and is 
  therefore equivalent to optimizing the precision.

\begin{table}[htbp]
\begin{tabular}{|c l|c|c|}
\hline \hline
  &               & $\nu$  & $\eta$    \\
\hline \hline
          &             & \hspace{0.45cm}  full    \hspace{0.42cm}      truncated  \hspace{0.1cm}  &   \hspace{0.45cm}  full    \hspace{0.42cm}      truncated  \hspace{0.1cm}   \\
\hline
 &LPA           & \hspace{0.2cm}  0.6503 \hspace{0.4cm}   0.6506  \hspace{0.2cm}   &  \hspace{0.2cm}  0  \hspace{0.4cm}     0   \hspace{0.2cm}   \\
(a) &FOA                    &  \hspace{0.2cm} 0.6291  \hspace{0.4cm}  0.6267     \hspace{0.2cm}     &  \hspace{0.2cm}  0.1058 \hspace{0.4cm}  0.1058   \hspace{0.2cm}    \\
 &SOA                      & \hspace{0.2cm}  0.6277  \hspace{0.4cm}  0.6281 \hspace{0.2cm}   &  \hspace{0.2cm} 0.0443  \hspace{0.4cm}  0.0443  \hspace{0.2cm}  \\
\hline \hline
(b)  &7-loop \;\;       & 0.6304(13)& 0.0335(25)\\ 
(c)  &MC            & 0.6297(5) & 0.0362(8)\\
\hline \hline
\end{tabular}
\caption{Critical exponents of the three-dimensional Ising model:
$a$) effective average action method: ``full'' stands for results keeping the full field-dependence
 of $u_k$ and $z_k$ (present work), ``truncated'' denotes results  from  field expansions of these functions 
 \cite{canet03a}; 
$b$) 6-loop calculations including 7-loop corrections \cite{guida98}; $c$) Monte-Carlo simulations
\cite{hasenbusch01}.}
\label{tab}
\end{table}

\begin{figure}[ht]
\begin{center}
\includegraphics[width=85mm,height=58mm,angle=-90]{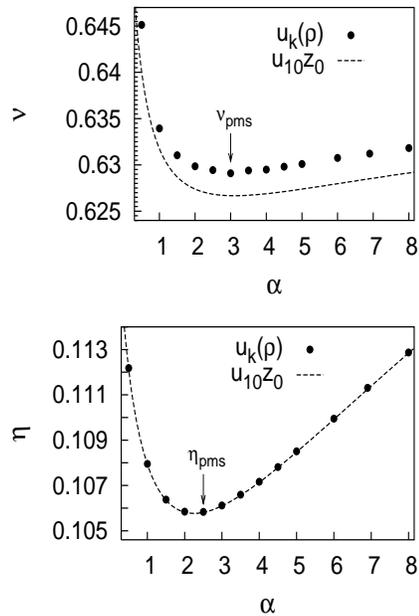}
\caption{Variations of $\nu$ and $\eta$ with $\alpha$ at the FOA.
 The dashed line represents results obtained 
 with a tenth order truncated potential and a field
renormalization coefficient \cite{canet03a}, dots those
 obtained keeping the full functional dependence of $u_k(\trho)$ (present work).} 
\label{pmsd1}
\end{center}
\end{figure}

 We then come to the FOA, for which the anomalous dimension
 becomes nontrivial. 
  The variations of $\nu$ and $\eta$ with the parameter $\alpha$ are displayed in Figure \ref{pmsd1},
 compared with the curves following from a  truncated potential  (at $p_u=10$)
 with a field renormalization coefficient $Z_{k,0}$ (which corresponds to
 $p_z=0$ in Eq. (\ref{fieldexp})) \cite{canet03a}.
 Both approaches yield very similar results. Let us note
  that, for both, the anomalous dimension is defined
  by the fixed point value of $\eta_k$ evaluated at the minimum of the potential 
    (see Eq. (\ref{defeta})). Hence the corresponding results are not expected
 to differ much. Indeed,  Figure \ref{pmsd1} shows that 
 $\eta(\alpha)$ computed from both procedures coincide. 
 On the other hand,  whether or not a field
truncation is implemented alters the determination of $\nu$. Only values of the potential and its
derivatives {\it at} the minimum
 enter the computation of $\nu$ when the latter is field expanded,
 whereas, when integrating the
field-dependent flows, the determination of this exponent
  mixes values on 
 the whole field mesh considered and thus incorporates a richer information.
  However, the $\nu(\alpha)$ curves from both procedures also reveal a good agreement,
  below the percent level (see Figure \ref{pmsd1} and Table \ref{tab}). Hence the smallness of the discrepancy 
   seems to indicate that the vicinity of the minimum already
  captures most of the relevant features. 

 We can now discuss the optimization at the FOA. Both $\nu(\alpha)$ and $\eta(\alpha)$ still exhibit a single extremum and thus a unique PMS solution:  $\nu_\pms=0.6291$ and $\eta_\pms = 0.1058$.
 Again, these exponents  turn out to minimize the distance to the best 
 theoretical estimates (see Table \ref{tab}), since the latters lie
  below all $\nu(\alpha)$ and  $\eta(\alpha)$ respectively and the PMS solutions are minima.
  The same conclusion can therefore 
 be reiterated as for the equivalence between the PMS and the optimization of the accuracy. 

\begin{figure}[ht]
\begin{center}
\includegraphics[width=85mm,height=58mm,angle=-90]{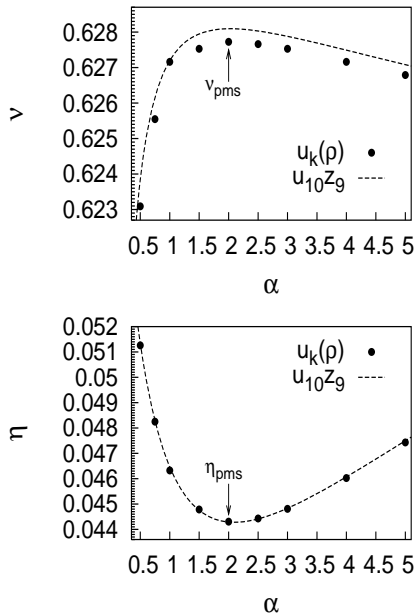}
\caption{Variations of $\nu$ and $\eta$ with $\alpha$ at the SOA.
  The dashed line represents results obtained 
 with a tenth order truncated potential  and a ninth order truncated field
renormalization function  \cite{canet03a}, dots those
 obtained keeping the full functional dependence  of both $u_k(\trho)$  and
$z_k(\trho)$ (present work).}
\label{pmsd2}
\end{center}
\end{figure}

 Let us finally investigate the SOA, considering the full field-dependence of the field renormalization function
 $z_k(\trho)$. This order involves a considerable numerical task \footnote{The obtention of 
a single point of Figure \ref{pmsd2} demands on average 35 days  CPU of a 3.2GHz  Pentium IV.}.
 Figure \ref{pmsd2} shows the variations of the two 
 critical exponents with $\alpha$, together with the analogous curves
obtained  by field expanding both  $u_k(\trho)$ and  $z_k(\trho)$
 (to $p_u=10$ and $p_z=9$  respectively) \cite{canet03a}. First, these curves support the previous discussion 
 as for the two procedures --- with and without truncation --- as the anomalous dimensions 
 exactly match, and the $\nu$ exponents lie within a percent.
 Moreover, at this order, the $\nu(\alpha)$ curvature is reversed such that the unique  
 extremum becomes a maximum. However, $\nu(\alpha)$ turns out to underestimate
 the  ``expected'' exponent for all $\alpha$, and thus the PMS solution still minimizes the distance to the best theoretical value.
 Indeed, $\nu_\pms = 0.6277$ appears in close agreement with the 7-loop estimate $\nu = 0.6307(13)$.
 On the other hand, $\eta(\alpha)$ always overvalues the  ``expected'' anomalous dimension and $\eta_\pms = 0.0443$  achieves the minimum. Thus, 
 the PMS  once again selects the most accurate exponents at a given truncation. \\

 We have  proposed an optimization procedure 
 applicable when dealing with  field-dependent NPRG flows.
 The PMS has indeed appeared as an efficient tool, allowing to optimize the
 accuracy of the exponents of the three-dimensional Ising model, and thus constitutes a useful mean
 of control of the NPRG formalism even when field  expansions are not possible or not
 desirable.  This procedure would be particularly valuable to confirm the results obtained 
 at the next order ($\p^4$) 
 of the derivative expansion \cite{canet03b}, 
 since verifying explicitely the convergence of the field expansion at this order becomes extremely tedious.
 Furthermore, such a procedure would ensure one to control the influence of $R_k$ and thus to work out  
    reliable results within physical contexts
  requiring a functional description --- for instance to deal with non-analytical potentials.\\

\acknowledgments
We wish to thank  B. Delamotte,  D. Mouhanna and J. Vidal
  for  fruitful discussions and careful readings.
 The LPTHE is  Unit\'{e} Mixte du CNRS  UMR 7589.


\end{document}